# Experimental Observation of Quantum Hall Effect and Berry's Phase in Graphene

Yuanbo Zhang[1], Yan-Wen Tan[1], Horst L. Stormer[1,2] & Philip Kim[1]

[1]*Department of Physics, Columbia University, New York, New York 10027, USA* [2]*Department of Applied Physics and Applied Mathematics, Columbia University, New York, New York 10027, USA*

**When electrons are confined in two-dimensional (2D) materials, quantum mechanically enhanced transport phenomena, as exemplified by the quantum Hall effects (QHE), can be observed. Graphene, an isolated single atomic layer of graphite, is an ideal realization of such a 2D system. Its behaviour is, however, expected to differ dramatically from the well-studied case of quantum wells in conventional semiconductor interfaces. This difference arises from the unique electronic properties of graphene, which exhibits electron-hole degeneracy and vanishing carrier mass near the point of charge neutrality.[1,2] Indeed, a distinctive half-integer QHE [3-5] has been predicted theoretically, as has been the existence of a non-zero Berry's phase of the electron wavefunction, a consequence of the exceptional topology of the graphene band structure.[6,7] Recent advances in micromechanical extraction and fabrication techniques for graphite structures[8-12] now permit such exotic 2D electron systems to be probed experimentally. Here, we report an experimental investigation of magneto transport in a high mobility single layer of graphene. Adjusting the chemical potential using the electric field effect, we observe an unusual half integer QHE for both electron and hole carriers in graphene. The relevance of Berry's phase to these experiments is confirmed by magneto-oscillations. In addition to their purely scientific interest, these unusual quantum transport phenomena suggest carbon-based novel electronic and magneto-electronic device applications.**

The low energy band structure of graphene can be approximated as cones located at two inequivalent Brillouin zone corners (left inset to Fig 1a). In these cones, the 2D energy dispersion relation is linear and the electron dynamics can be treated as 'relativistic,' where the Fermi velocity $v_F$ of the graphene plays the role of the speed of light. In particular, at the apex of the cones (termed Dirac point), electrons and holes (particles and antiparticles) are degenerate. Landau level (LL) formation for electrons in such system under a perpendicular magnetic field, *B*, has been studied theoretically using an analogy to the 2+1 dimensional Quantum Electro Dynamics (QED)[2,3], where the Landau level energy is given by

$$E_n = \text{sgn}(n)\sqrt{2e\hbar v_F^2 |n| B} \qquad (1)$$

Here $e$ and $\hbar = h/2\pi$ are electron charge and Plank's constant divided by $2\pi$, and the integer *n* represents an electron-like ($n >0$) or a hole-like ($n <0$) LL index. Crucially, a single LL with $n=0$ and $E_0=0$ also occurs. When only low lying LLs ($|n|<10^4$ for $B=10$ T) are occupied, the separation of $E_n$ is much larger than the Zeeman spin splitting, thus each LL has a degeneracy $g_s=4$, accounting for spin degeneracy and sublattice degeneracy. Previous studies of mesoscopic graphite samples consisting of a few layers of graphene exhibited magneto-oscillations associated with the LL formation by electron-like and hole-like carriers tuned by the electric field effect [8, 9, 11]. However, the QHE was not observed in these samples, possibly due to their low mobility and/or the residual 3D nature of the specimens.

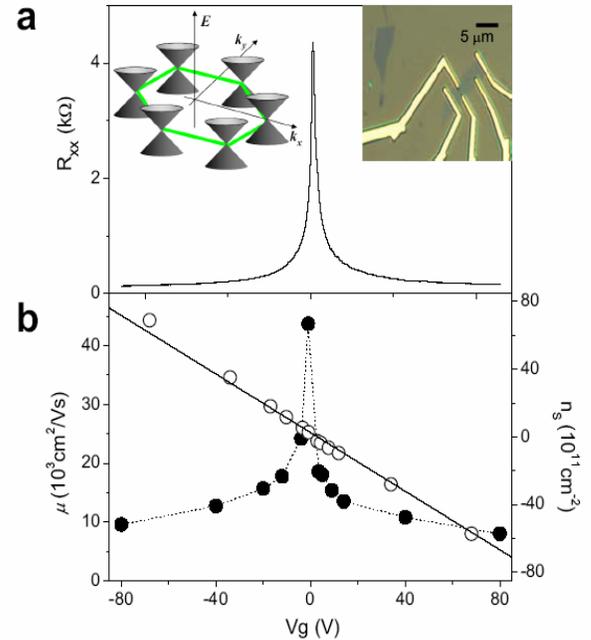

**Figure 1** Resistance, carrier density, and mobility of graphene measured at 1.7 K at different gate voltages. **a,** Resistance changes as a function of gate voltage in a graphene device shown in the optical microscope image in the right inset. The position of the resistance peaks varies from device to device, but the peak values are always of the order of ~4kΩ, suggesting a potential quantum mechanical origin. The left inset shows a schematic diagram for the low energy dispersion relation near the Dirac points in the graphene Brillouin zone. Only two Dirac cones are inequivalent to each other, producing a two-fold valley degeneracy in the band structure. **b,** Charge carrier density (open circle) and mobility (filled circle) of graphene as a function of gate voltage. The solid line corresponds to the estimated charge induced by the gate voltage, $n_s = C_g V_g / e$, assuming a gate capacitance $C_g$ = 115 aF/μm$^2$ obtained from geometrical consideration.

The high mobility graphene samples used in our experiments were extracted from Kish graphite (Toshiba Ceramics Co.) on degenerately doped Si wafers with a 300 nm $SiO_2$ coating layer, using the micromechanical manipulation similar to that described in ref. [8]. Interference induced colour shifts, cross-correlated with an atomic force microscopy profile, allow us to identify the number of deposited graphene layers from optical images of the samples. After a suitable graphene sample is selected, electron beam lithography followed by thermally evaporated Au/Cr (30/5 nm) defines multiple electrodes for transport measurement (right inset to Fig 1a). Using a Hall-bar type electrode configuration, the magnetoresistance $R_{xx}$ and Hall resistance $R_{xy}$ are measured. Applying a gate voltage, $V_g$, to the Si substrate controls the charge density in the graphene samples.

Fig 1a shows the gate modulation of $R_{xx}$ at zero magnetic field in a typical graphene device whose lateral size is ~ 3 μm. While $R_{xx}$ remains in the ~ 100 Ω range at high carrier density, a sharp peak whose value is on the order of ~ 4 kΩ is observed at $V_g \approx 0$. While different samples show slightly different peak values and peak position, similar behaviours were observed in 3 other graphene samples we have measured. The existence of such a sharp peak is consistent with the reduced carrier density as $E_F$ approaches the Dirac point of graphene where the density of states vanishes. Thus the gate voltage corresponding to the charge neutral Dirac point, $V_{Dirac}$, can be determined from this peak position. A separate Hall measurement provides a measure for the sheet carrier density, $n_s$, and for the mobility, $\mu$, of the sample, as shown in Fig 1b, assuming a simple Drude model. The sign of $n_s$ changes at $V_g = V_{Dirac}$, indicating $E_F$ indeed crosses the charge neutral point. Mobilities are higher than $10^4$ $cm^2/Vs$ for the entire gate voltage range, considerably exceeding the quality of previously studied graphene samples[8, 9].

The exceptionally high mobility graphene samples allow us to investigate transport phenomena in the extreme magnetic quantum limit, such as the QHE. Fig. 2a shows $R_{xy}$ and $R_{xx}$ of the sample of Fig1 as a function of magnetic field $B$ at a fixed gate voltage $V_g>V_{Dirac}$. The overall positive $R_{xy}$ indicates that the contribution is mainly from electrons. At high magnetic field, $R_{xy}(B)$ exhibits plateaus and $R_{xx}$ is vanishing, which are the hallmark of the QHE. At least two well-defined plateaus with values $(2e^2/h)^{-1}$ and $(6e^2/h)^{-1}$, followed by a developing $(10e^2/h)^{-1}$ plateau, are observed before the QHE features transform into Shubnikov de Haas (SdH) oscillations at lower magnetic field. The quantization of $R_{xy}$ for these first two plateaus is better than 1 part in $10^4$, precise within the instrumental uncertainty. We observed the equivalent QHE features for holes ($V_g<V_{Dirac}$) with negative $R_{xy}$ values (Fig 2a, inset). Alternatively, we can probe the QHE in both electrons and holes by fixing the magnetic field and changing $V_g$ across the Dirac point. In this case, as $V_g$ increases, first holes ($V_g<V_{Dirac}$) and later electrons ($V_g>V_{Dirac}$) are filling successive Landau levels and exhibit the QHE. This yields an antisymmetric (symmetric) pattern of $R_{xy}$ ($R_{xx}$) in Fig. 2b, with $R_{xy}$ quantization accordance to

$$R_{xy}^{-1} = \pm g_s (n+\frac{1}{2})\frac{e^2}{h} \qquad (2)$$

where $n$ is a non-negative integer, +/- stands for electrons and holes respectively. This quantization condition can be translated to the quantized filling factor $\nu = \pm g_s(n+1/2)$ in the usual QHE language. In addition, there is an oscillatory structure developed near the Dirac point. Although this structure is reproducible for any given sample, its shape varies from device to device, suggesting potentially mesoscopic effects depending on the details of the sample geometry[13]. While the QHE has been observed in many 2D systems, the QHE observed in graphene is distinctively different those 'conventional' QHEs since the quantization condition (eq (2)) is shifted by a half integer. These unusual quantization conditions are a result of the topologically exceptional electronic structure of graphene which we discuss below.

The sequence of half-integer multiples of quantum Hall plateaus has been predicted by several theories which combine 'relativistic' Landau levels with the particle-hole symmetry of graphene[3-5]. This can be easily understood from the calculated LL spectrum (eq (1)) as shown in Fig. 2c. Here we plot the density of states (DOS) of the $g_s$-fold degenerate (spin and sublattice) of LLs and the corresponding Hall conductance ($\sigma_{xy}=-R_{xy}^{-1}$, for $R_{xx} \to 0$) in the quantum Hall regime as a function of energy. $\sigma_{xy}$ exhibits QHE plateaus when $E_F$ (tuned by $V_g$) falls between LLs, and jumps by an amount of $g_s e^2/h$ when $E_F$ crosses a LL. Time reversal invariance guarantees particle-hole symmetry and thus $\sigma_{xy}$ is an odd function in energy across the Dirac point[2]. However, in graphene, the $n=0$ LL is robust, i.e., $E_0=0$ regardless of the magnetic field, provided that the sublattice symmetry is preserved[2]. Thus the first plateau of $R_{xy}^{-1}$ for electron and hole are situated exactly at $\pm g_s e^2/2h$. As $E_F$ crosses the next electron (hole) LL, $R_{xy}^{-1}$ increases (decreases) by an amount of $g_s e^2/h$, which yields the quantization condition in eq. (2).

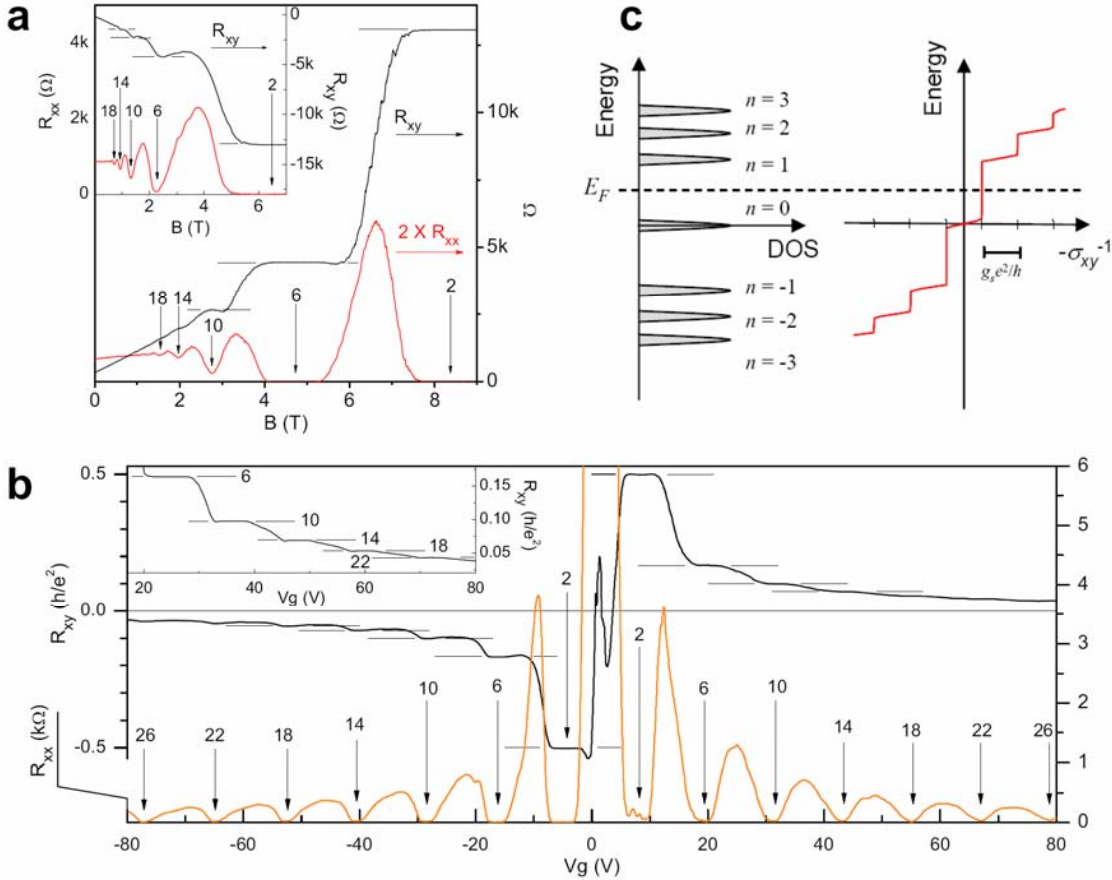

**Figure 2** Quantized magnetoresistance and Hall resistance of a graphene device. **a,** Hall resistance (black) and magnetoresistance (red) measured in the device in Fig. 1 at $T = 30$ mK and $V_g = 15$ V. The vertical arrows and the numbers on them indicate the values of $B$ and the corresponding filling factor $\nu$ of the quantum Hall states. The horizontal lines correspond to $h/e^2 \nu$ values. The QHE in the electron gas is demonstrated by at least two quantized plateaus in $R_{xy}$ with vanishing $R_{xx}$ in the corresponding magnetic field regime. The inset shows the QHE for a hole gas at $V_g = -4$ V, measured at 1.6 K. The quantized plateau for filling factor $\nu = 2$ is well-defined and the second and the third plateau with $\nu = 6$ and 10 are also resolved. **b,** The Hall resistance (black) and magnetoresistance (orange) as a function of gate voltage at fixed magnetic field $B = 9$ T, measured at 1.6 K. The same convention as in **a** is used here. The upper inset shows a detailed view of high filling factor plateaus measured at 30 mK. **c,** A schematic diagram of the Landau level density of states (DOS) and corresponding quantum Hall conductance ($\sigma_{xy}$) as a function of energy. Note that in the quantum Hall sates, $\sigma_{xy}=-R_{xy}^{-1}$. The LL index $n$ is shown next to the DOS peak. In our experiment, the Fermi energy $E_F$ can be adjusted by the gate voltage, and $R_{xy}^{-1}$ changes by an amount of $g_s e^2/h$ as $E_F$ crosses a LL.

As noted by several workers, a consequence of the combination of time reversal symmetry with the novel Dirac point structure can be viewed in terms of Berry's phase arising from the band degeneracy point[7,14]. A direct implication of Berry's phase in graphene is discussed in the context of the quantum phase of a spin-1/2 pseudo-spinor that describes the sublattice symmetry [6,15]. This phase is already implicit in the half-integer shifted quantization rules of the QHE. It can further be probed in the magnetic field regime where a semi-classical magneto-oscillation description holds[16,17]:

$$\Delta R_{xx} = R(B,T) \cos\left[2\pi\left(\frac{B_F}{B} + \frac{1}{2} + \beta\right)\right] \quad (3)$$

Here $R(B,T)$ is the SdH oscillation amplitude, $B_F$ is the frequency of the SdH oscillation in $1/B$, and $\beta$ is the associated Berry's phase of value $0<\beta<1$. Berry's phase $\beta=0$ (or equivalently $\beta=1$) corresponds to the trivial case. A deviation from this value is indicative of interesting new physics with $\beta=½$ implying the existence of Dirac particles[7]. Experimentally, this phase shift in the semi-classical regime can be obtained from an analysis of the SdH fan diagram, in which the sequence of values of $1/B_n$ of the $n_{th}$ minimum in $R_{xx}$ are plotted against their index $n$ (Fig 3b). The intercept of linear fit to the data with the $n$-index axis yields Berry's phase, modulo an integer. Remarkably, the resulting $\beta$ is very close to 0.5 (upper inset to Fig 3b), providing further manifestation for the existence of a non-zero Berry's phase in graphene and the presence of Dirac particles. Such a non-zero Berry's phase was not observed in the previous few layer graphite specimens[8,11,18], although there have been claims of hints of a phase shift in earlier measurements on *bulk* graphite[17]. Our data for graphene provide indisputable evidence for such an effect in a solid state system.

The non-zero Berry's phase observed in the SdH fan diagram is related to the vanishing mass at the Dirac point. We can extract this effective carrier mass $m_c$ from the temperature dependence of the well developed SdH oscillations at low B-field (Fig 3a left inset) using the standard SdH formalism[19]. Indeed, the analysis at different gate voltages yields a strong suppression of $m_c$ near the Dirac point. While the high density $(n_s \sim 5\times10^{12}$ cm$^{-2})$ carrier gas shows $m_c \sim 0.04\ m_e$, the mass drops to $m_c \sim 0.007\ m_e$ near the Dirac point ($n_s \sim 2 \times 10^{11}$ cm$^{-2}$), where $m_e$ is the free electron mass. Overall, the observed gate voltage-dependent effective mass can be fit to an fictitious 'relativistic' mass: $m_c = E_F / v_F^2 = \sqrt{\pi \hbar^2 n_s} / v_F^2$ using $v_F$ as the only fitting parameter (Fig 3a right inset). In accordance with the Berry's phase argument, this procedure extrapolates to a vanishing mass at the Dirac point.

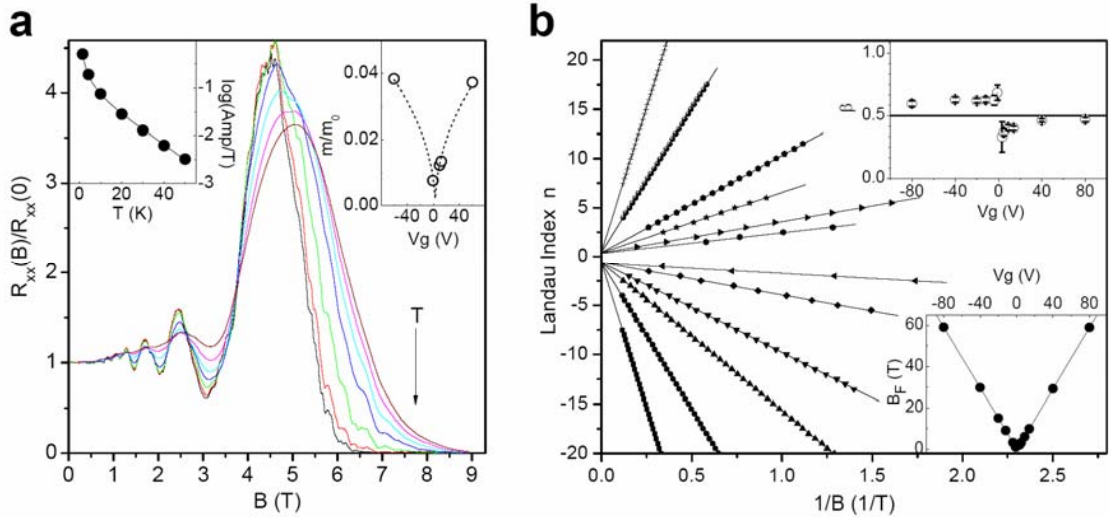

**Figure 3** Temperature dependence and gate voltage dependence of the Shubnikov de Haas oscillations in graphene. **a,** Temperature dependence of the SdH oscillations at $V_g = -2.5$ V. Each curve represents $R_{xx}(B)$ normalized to $R_{xx}(0)$ at a fixed temperature. The curves are in order of decreasing temperature starting from the top as indicated by the vertical arrow. The corresponding temperatures are listed in the left inset. The left inset represents the SdH oscillation amplitude divided by temperature measured at a fixed magnetic field. The standard SdH fit yields the effective mass. The right inset is a plot of the effective mass obtained at different gate voltages. The broken line is a fit to the single parameter model described in the text, which yields $v_F = 1.1\times10^6$ m/s, in reasonable agreement with the literature values. **b,** A fan diagram for SdH oscillations at different gate voltages. The location of $1/B$ for the $n_{th}$ minimum (maximum) of $R_{xx}$ counting from $B = B_F$ is plotted against $n$ $(n+1/2)$. The lines correspond to a linear fit, where the slope (lower inset) indicates $B_F$ and the $n$-axis intercept (upper inset) provides a direct probe of Berry's phase in the magneto-oscillation in graphene.

In conclusion, we have experimentally discovered an unusual QHE in high quality graphene samples. Different from conventional 2D systems, in graphene, the observed quantization condition is described by half integer rather than integer values. The measured phase shift in magneto-oscillation can be attributed to the peculiar topology of the graphene band structure with a linear dispersion relation and vanishing mass near the Dirac point, which can be described in terms of fictitious 'relativistic' carriers. The unique behaviour of electrons in this newly discovered 2+1 dimensional QED system not only opens up many interesting questions in mesoscopic transport in electronic systems with non-zero Berry's phase but may also provides the basis for novel carbon based electric and magnetic field effect device applications, such as ballistic metallic/semiconducting graphene ribbon devices[9] and electric field effective spin transport devices utilizing spin-polarized edge state[20].

**Acknowledgements** We thank I. Aleiner, T. Heinz, A. Millis, A. Mitra, J. Small and A. Geim for discussion. This research was supported by the NSF Nanoscale Science and Engineering Center at Columbia University, New York State Office of Science (NYSTAR), and Department of Energy (DOE).

Correspondence and requests for materials should be addressed P.K. (e-mail: pkim@phys.columbia.edu).